\documentclass[journal=cmatex]{achemso}

\def\@xfootnote[#1]{%
	\protected@xdef\@thefnmark{#1}%
	\@footnotemark\@footnotetext}
\makeatother

\usepackage{amsmath} 
\usepackage{amssymb}
\newcommand{\angstrom}{\text{\normalfont\AA}}

\usepackage[percent]{overpic}
\usepackage{wrapfig}
\usepackage{xcolor}
\usepackage{lipsum}
\usepackage{graphicx}
\usepackage{epstopdf}
\usepackage{dcolumn}
\usepackage{bm}
\usepackage{bigints}
\usepackage{supertabular}
\usepackage{caption}
\usepackage{pdfpages}

\graphicspath{ {./fig/} }

	\title[Ultra Low Cobalt Cathodes]{Towards Ultra Low Cobalt Cathodes: A High Fidelity Computational Phase Search of Layered Li-Ni-Mn-Co Oxides}
	
	\author{Gregory Houchins}
	\affiliation{Department of Physics, Carnegie Mellon University, Pittsburgh, Pennsylvania 15213, USA}
	\affiliation{Wilton E. Scott Institute for Energy Innovation, Carnegie Mellon University, Pittsburgh, Pennsylvania 15213, USA}
	\author{Venkatasubramanian Viswanathan}%
	\email{venkvis@cmu.edu}
	\affiliation{Department of Physics, Carnegie Mellon University, Pittsburgh, Pennsylvania 15213, USA}%
	\affiliation{Department of Mechanical Engineering, Carnegie Mellon University, Pittsburgh, Pennsylvania 15213, USA}
	\affiliation{Wilton E. Scott Institute for Energy Innovation, Carnegie Mellon University, Pittsburgh, Pennsylvania 15213, USA}
	\date{\today}

\begin{document}

	\begin{abstract}
	    Layered Li(Ni,Mn,Co,)O$_2$ (NMC) presents an intriguing ternary alloy design space for optimization as a cathode material in Li-ion batteries. In the case of NMC, however, only a select few proportions of transition metal cations have been attempted and even fewer have been adopted on a large scale. Recently, the high cost and resource limitations of Co have added a new design constraint and high Ni-containing NMC alloys have gained enormous attention despite possible performance trade-offs. Although the limited collection of NMC cathodes have been successful in providing the performance needed for many applications, specifically electric vehicles, this concern around Co requires further advancement and optimization within the NMC design space. Additionally, it is not fully understood if this material space is a disordered solid solution at room temperature and any arbitrary combination can be used or if there exist distinct transition metal orderings to which meta-stable solid solutions will decay during cycling and affect performance. Here, we present a high fidelity computational search of the ternary phase diagram with an emphasis on high-Ni, and thus low Co, containing compositional phases to understand the room temperature stability of the ordered and disordered solid solution phases. This is done through the use of density functional theory training data fed into a reduced order model Hamiltonian that accounts for effective electronic and spin interactions of neighboring transition metal atoms at various lengths in a background of fixed composition and position lithium and oxygen atoms. This model can then be solved to include finite temperature thermodynamics into a convex hull analysis to understand the regions of ordered and disordered solid solution as well the transition metal orderings within the ordered region of the phase diagram.  We also provide a method to propagate the uncertainty at every level of the analysis to the final prediction of thermodynamically favorable compositional phases thus providing a quantitative measure of confidence for each prediction made.  Due to the complexity of the three component system, as well as the intrinsic error of density functional theory, we argue that this propagation of uncertainty, particularly the uncertainty due to exchange-correlation functional choice is necessary to have reliable and interpretable results. We find that for the majority of transition metal compositions of the layered material, specifically medium to high-Ni content, prefer transition metal ordering and predict the collection of preferred compositions in the ordered region. 

	\end{abstract}
	
	\maketitle
	

	\section{\label{sec:intro}Introduction}

	The first rechargeable Li-ion battery cathode material, layered LiCoO$_2$,\cite{Mizushima1980} revolutionized the world of portable electronics. The high cost and low operational capacity of this battery material was improved with the addition of Ni and Mn to create LiNi$_x$Mn$_y$Co$_{(1-x-y)}$O$_2$ (NMC). This was first successfully done by Lui et al.\cite{Liu1999} and later the most widespread phase, x=y=1/3 (NMC111) was synthesized by Ohzuku et al.\cite{Ohzuku2001} Since then a small collection of phases have been used in various applications with some of the most promising candidates being x=0.5, y=0.3 (NMC532) \cite{Wang2003,NMC532,Li2017}, and x=0.8,y=0.1 (NMC811) \cite{Choi2006,Kim2006}. Additionally, NMC cathode materials are now present in a majority of electric vehicle battery chemistries \cite{Blomgren2017}.
	
    Many studies have explored trends of operational performance as a function of Ni, Mn, and Co content. Julien et al. demonstrated that with increasing Co content, the specific capacity increases\cite{Julien2016}. Manthiram et al  on the other hand demonstrated a capacity increase with increasing Ni content \cite{Manthiram2016}. The discrepancy between these studies can be attributed to their distinct, one-dimensional sweeps of the full two-dimensional phase space. Another consideration regarding transition metal composition is that of thermal stability. Recent work has shown that thermal stability decreases with increasing Ni content\cite{Bak2014}. In this study, however, no phases were tested for Ni content between 60$\%$ and 80$\%$ and only one possible combination of Co and Mn content was used for each Ni content. Only with a full understanding of the entire compositional phase space, can trends such as these be understood and possibly broken.  Additionally within all of these studies, the material space is assumed to be a disordered solid solution. While this is likely true at synthesis temperatures given the synthesizability of NMC at arbitrary compositions, this is likely not true at ambient temperatures at which the material is electrochemically cycled within a battery. In fact a tendency for ordering has been seen even for NMC materials that were rapidly quenched after synthesis \cite{Zeng2007,Cahill2005, Yabuuchi2005, Koyama2003, Yoon2004}. Additionally an influence of this ordering on aspects such as the voltage profile during cycling has also been seen \cite{Cahill2005}. Thus the proper ordering of the transition metals is necessary for proper simulation of the material. Many previous computational works have provided reasonable guesses for various other phases with, however, no exploration for the stability with respect to other cation orderings at the same composition \cite{NMC532,Liang2017,Xu2017}.  Within this work, we also specifically attempt to address the need for new low-Co content cathode materials as Co production presents major price volatility and supply concerns in the near future. \cite{Olivetti2017}  Amid these concerns of material availability, the United States Department of Energy has identified the need for low-Co cathode materials as imperative for future battery technologies, setting a target of 50 $\frac{mg}{Wh}$ of Co \cite{doe}. This target however is far from realization in materials that can meet the performance needs currently satisfied by high-Co cathodes. We attempt to understand if the region of ultra low Co is a disordered solid solution where any composition can be expected to have stable transition metal orderings, or if there are preferred compositions and orderings that will emerge within low Co cathodes ultimately cause inhomogeneous electrochemical activity and cycling. 
    

    Density functional theory (DFT) alongside thermodynamic modeling has been successfully used as a method for understanding and predicting the stable phases of many binary alloys \cite{vandeWalle2002,Ruban2008} and battery materials \cite{VanderVen2004, Wolverton1998, VanderVen1998}. Density functional theory calculations train a computationally efficient model and with the use of statistical simulations using the model Hamiltonian, the thermodynamically stable phases with respect to composition can be predicted. 
    
    One of the largest factors in the accuracy of density functional theory calculations is the choice of the exchange correlation functional. Different exchange correlation functionals are shown to have drastically different levels of accuracy for different material classes with certain functionals performing better for certain applications \cite{Peverati2014,Mardirossian2017}. Recently, a class of functionals known as Bayesian Error Estimation Functionals have been developed to incorporate a collection of exchange correlation functionals and therefore provide an estimate of uncertainty related to the calculation of exchange correlation energetics. Previous work has applied this uncertainty estimation to the prediction of magnetic ground states, bulk properties, and activity of catalysts \cite{c-value,ZeeshanPRB,DilipJPC,Deshpande2016}.

    Given the complexity of the problem, it is necessary to fully understand the uncertainty from systematic errors in the DFT training data as well as random errors from the statistical nature of Monte Carlo simulations. We focus most heavily on understanding the uncertainty derived from density functional theory. Reduced order models such as cluster expansion can be fit to high accuracy given enough terms, and the statistical error of Monte Carlo can be controlled and understood given a sufficient number of sweeps in your simulation as well as repeated simulations. The systematic error in density functional theory, however, is impossible to control as there is no way to find an exchange correlation functional for an arbitrary system with arbitrary accuracy. We therefore attempt to understand the sensitivity of our results to variation of the exchange correlation functional within a systematically chosen subspace of exchange-correlation functional space. Additionally, we attempt to show that without considering uncertainty propagation, the goal of identifying promising new low-Co phases could not be done with trustable results. We do however find that for the majority of the computational phase space, the energetic difference between the ordered and disorder phase is much larger than the the uncertainty of our calculations. Ultimately, within this work, we present an extensive computational search of the ternary phase diagram of NMC cathodes to understand the compositional regions of order versus disorder, as well as the most favorable compositions in the ordered region. 
     
    \begin{figure}
		\includegraphics[width=.48\textwidth]{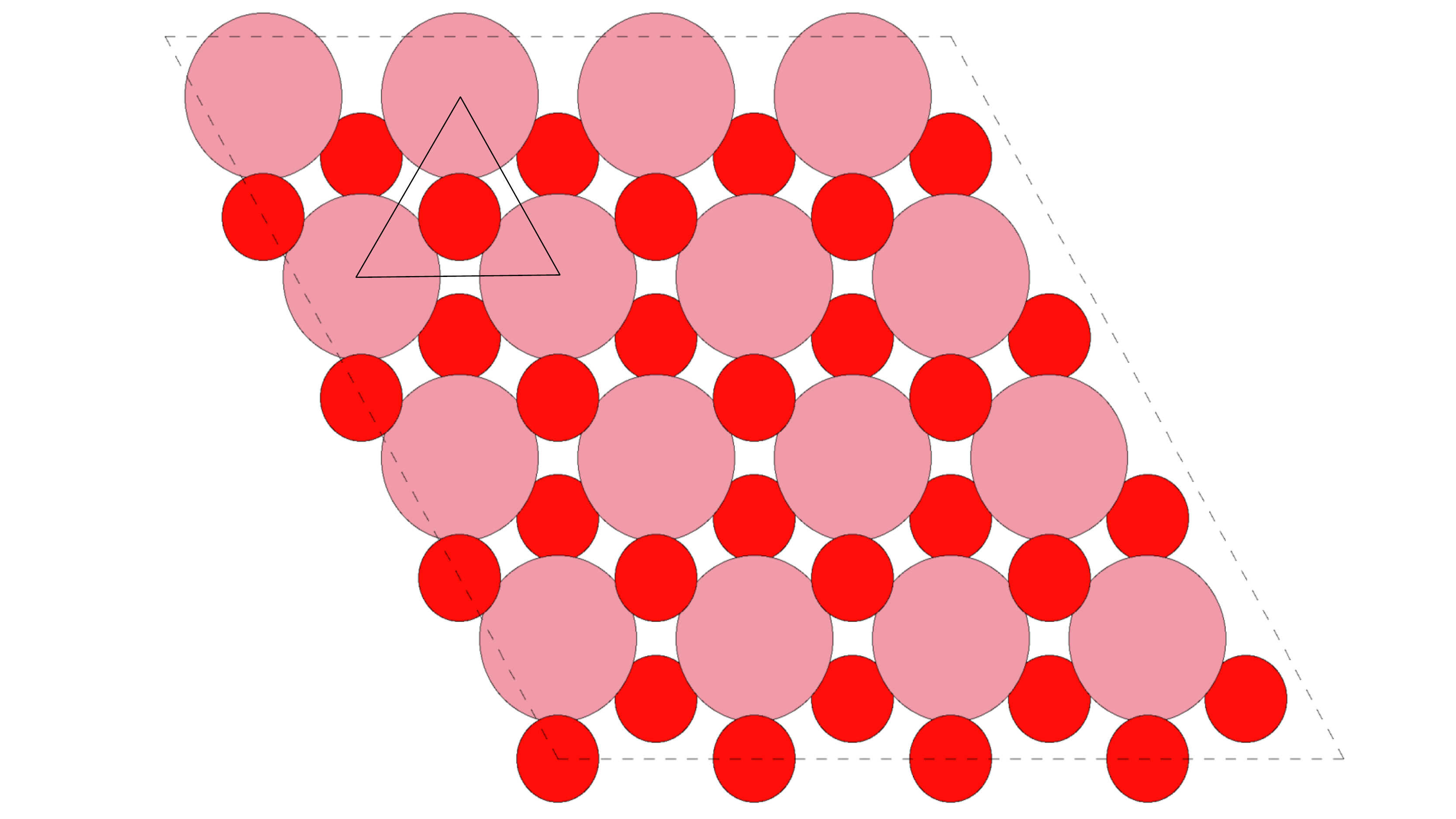}
		\caption{\label{latttice} A top view of the lattice structure. The cation site are shown pink, while the oxygen sites are red and the lithium atoms are not pictured. The black triangle shows the idealized triangle lattice the cations form.}
	\end{figure}

    \section{Computational Details}  
	    In this work, we use a total of 118 density functional theory calculations for the formation energy of various NMC compositions, focusing on high Ni-content structures, to train a reduced order model in order to more efficiently evaluate the formation energy with only a minor decrease in accuracy compared to input. The reduced order model is then solved in a high fidelity search of the composition space using Monte Carlo simulations that incorporate finite temperature effects. From these simulations, the change in Gibbs free energy at 298 K from pure layered metal oxide end member states to a NMC compositions at a given composition is fed into a convex hull analysis where we are able to predict all thermodynamically stable NMC phases. Here we describe briefly each step of this process providing the full details in the Supplementary Materials.

    For all of the 118 density functional calculations used as training data, the Bayesian Error Estimation Functional with van der Waals (BEEF-vdW) \cite{BEEF} was used to treat the exchange-correlation energy at the level of the generalized gradient approximation. Previous work has shown that systematic error related to oxide materials can be reduced if the prediction of energy differences is made with the proper reference state without the use of Hubbard U, even in the case of transition metal oxides where a U may commonly be used \cite{Martinez2009,Rune2015JPC}. This was done by referencing the oxygen energy in the oxide to water rather than $O_2$ in order to match the oxidation state. This idea was further extended to show that the error in DFT predicted formation enthalpies of two similar reactions are correlated. That is the difference in formation enthalpies is constant and independent of exchange correlation function allowing the difference in these formation enthalpies to be compared accurately without the use of a Hubbard U correction \cite{Rune2016}. Again this is achieved by comparing two similar chemical environments to one another. Within this work, we therefore chose to predict and compare the formation energy of the mixed NMC phases with respect to the pure layered oxide end members and do not include the Hubbard correction within our calculations. As this reference compares oxygens in nearly identical chemical environments, we expect the errors to be reduced drastically versus the prediction of the formation energy from elements. 
    
    The BEEF-vdW functional has the ability to estimate the error of density functional theory with respect to experiment through the generation of an ensemble of energetic predictions and was chosen for this reason. The ensemble of energies is generated by feeding the electron density of a fully self consistent calculation of the empirically fit functional from BEEF-vdW to an ensemble of exchange correlation functionals. The spread of the energetic predictions has been pretuned to recreate the error of the DFT calculation with respect to the experimental data it was trained on. In this way, BEEF-vdW links the precision of the ensemble of predictions, to the accuracy of the prediction of the self consistent calculations. With this empirical error estimation, we can provide a quantification of uncertainty within each DFT calculation and propagate that uncertainty to the model parameters that are trained with respect to the DFT input. Beyond the result that the spread of energies estimates the uncertainty of the calculation, the ensemble of functionals provides a way to probe exchange-correlation space to understand the confidence of a prediction at the level of the generalized gradient approximations. This aspect can be used to ask if another exchange-correlation functional was used, would the prediction qualitatively change. The spread of the BEEF functional estimates has been been shown to bound the prediction of other general gradient approximation functionals for mechanical properties, magnetic ground states including, and reaction enthalpies for hydrocarbons \cite{ZeeshanPRB,c-value,Rune2015CST,Rune2016}. As there is no way to ensure accuracy or even the estimation of accuracy, we argue this aspect of uncertainty quantification strengthens the interpretability of DFT prediction and predictions made by models trained on DFT.

	Once the DFT data is generated, a reduced order model for the formation enthalpy with respect to the layered transition metal oxide end-members is chosen. This reference is chosen as it has been shown that for Ni, Mn, and Co can all be synthesized in the layered structure \cite{Mizushima1980,Ohzuku1993,Armstrong1996}. The change in volume of the lattice from pure states to the mixed state simulated by DFT is on the order of 1$\angstrom^3$ and therefore the enthalpy is assumed to be the same as the energy predicted from DFT. The goal of this reduced order model is to recreate the change in enthalpy with respect to the end members of homogeneous layered lithium metal oxides, rather than the enthalpy from constituent elements. Therefore, our model focuses on the effective interactions of the changing compositions and arrangement of the transition metal ions. The energetic effects of these constant background lithium and oxygen atoms with each other and a particular cation should be largely subtracted away as these interaction are assumed to be independent of composition. Additionally, we explicitly choose to keep the composition of lithium fixed at the maximum amount of lithium stoichiometrically possible so that there are no lithium vacancies and therefore no lithium-vacancy orderings to consider. When considering our results in the context of an operational cathode, we therefore assume the transition metal orderings predicted here to be relative constant over cycling. By treating the lithium and oxygen atoms as a constant background, that is only considering them indirectly within the model, we create a model lattice containing the relevant cations and only consider the energetics and interactions related to the proportions and orderings of the transition metal ions. This type of consideration of only specific interaction terms while leaving other atoms as a constant background term has been used successfully in other cluster expansion studies of transition metal oxide materials \cite{Fei2006,Malik2009,Lee2017}. We assume a triangle lattice where at each site on the lattice, one of the three metal cations (Ni, Mn, Co) is present as seen in Figure \ref{latttice}. The model used here, as with other cluster expansion models, contains energetic terms for the species that occupies a lattice site, and N-body interactions over various length scales as well as two body spin terms over various length scales given by: 
    \[\Delta H_f = \sum_{i,x}^{N} h_x \sigma_{i,x} + \sum_{\langle i,j \rangle ,x,y} J^{xy}_2 \sigma_{i,x} \sigma_{j,y} 
    + \sum_{\langle i,j,k \rangle ,x,y,z} J^{xyz}_3 \sigma_{i,x} \sigma_{j,y} \sigma_{k,z}+ \dots + \sum_{\langle i,j \rangle ,x,y} K_{xy} \vec{S}_{i,x} \cdot \vec{S}_{j,y} \]

	Where the indices i, j, k are sums over the lattice sites, $\langle i,j\rangle$ is a sum over two body interactions, and $\langle i,j, k \rangle$ a sum over three body interaction. Additionally, x, y, and z are sums over the three possible species (Ni,Mn,Co) and the J$_i$ terms represent the effective energetic strength of the particular cluster interaction involving the particular elements. Similar terms can be created in this way for larger n-body interactions for any arbitrary number. For the spin interaction terms, the $\vec{S}$ is the spin of the transition metal, and K is the interaction strength of the Ising like spin interaction in which the spins are assumed to be collinear.  
	
\begin{figure}
		\includegraphics[width=1\textwidth]{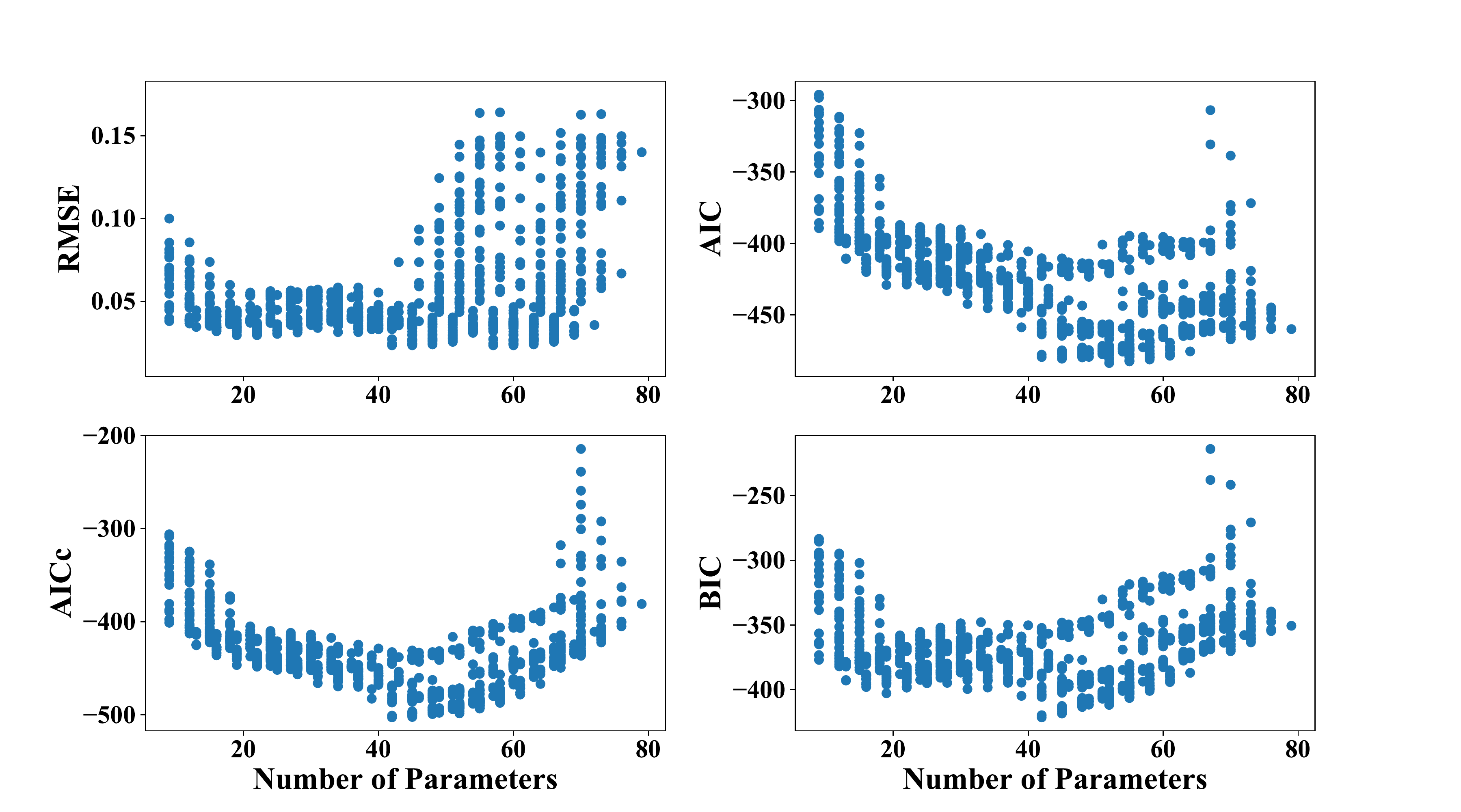}
		\caption{\label{model_selection} Plots for the root mean squared error (RMSE) from leave one out cross validation, the Akaike information criterion (AIC), the the corrected Akaike information criterion (AICc), and the Bayesian information critereon (BIC). }
	\end{figure}

	Within the full possible model space, which includes all heterogeneous 2, 3 and 4-body terms up to 6 $\angstrom$, and all 2-body spin terms within the transition metal layer up to 6 $\angstrom$, we then iterate over all possible collections of interactions and train that model on the data. It should be noted that for the DFT calculation, the Co atom converged to the a very small spin. The inclusion of Co spin interaction versus ignoring Co spin interactions was explored. In order to do this as well as select which of the interactions considered gave the best reduced order model, four model selection techniques were implemented. The goal of this model selection was to determine the best model that balances predictive power with model complexity to prevent over-fitting. Figure \ref{model_selection} shows the results of these tests as a function of number of terms in the model. 
	
	The first method used was leave one out cross validation where the model was trained on N-1 data points and the error was computed for the remaining data point. This was done N times so that each point was left out once and the root mean square of all of the errors was computed. Additionally the Akaike information criterion (AIC) \cite{Akaike1974} defined as 
	\[AIC = 2k - 2\ln{\hat{L}}  \]
	the corrected Akaike information (AICc) criterion\cite{Hurvich1989} defined as 
	\[AICc = AIC + \dfrac{2(k+1)(k+2)}{N-k-2} \]
	and the Bayesian information criterion (BIC) \cite{Schwarz1978} defined as 
	\[BIC = \ln{(N)} k - 2\ln{\hat{L}}  \]
	were also used. In each of the formulas, $k$ is the number of parameters in the model, $\hat{L}$ is the maximized likelihood function, and $N$ is the number of training points. We assume a Gaussian likelihood function for distribution of residual errors, $x_i$, and therefore the maximum likelihood estimators for the mean, $\mu$, and variance, $\sigma^2$, are the mean absolute error and the sample standard deviation given by

	\[\mu = \dfrac{\sum{x_i}}{N} \]
	\[\sigma^2= \dfrac{\sum{(x_i-\mu)^2}}{N}\]
     
     For each of these model selection techniques, the model with the lowest value of the metric is predicted to be the most ideal model. In our case, the same 42 parameter model which includes occupation, nearest neighbor spin not including Co spin, 4-body in plane, and 2-body out of plane next nearest neighbor interactions, was selected by leave one out cross validation, AICc, and BIC. The AIC selected a model that was slightly more complex than the 42 parameter model used which is a known bias of the AIC for smaller number of training points. \cite{Hurvich1989} The final model gave a root mean squared error from leave one out cross validation of 5.86 meV/atom.

\begin{figure}
	    \vspace{.5cm}
		\begin{minipage}[t]{0.475\linewidth}
			\begin{overpic}[width=1.0\linewidth,trim={0 0cm 0 0cm},clip]{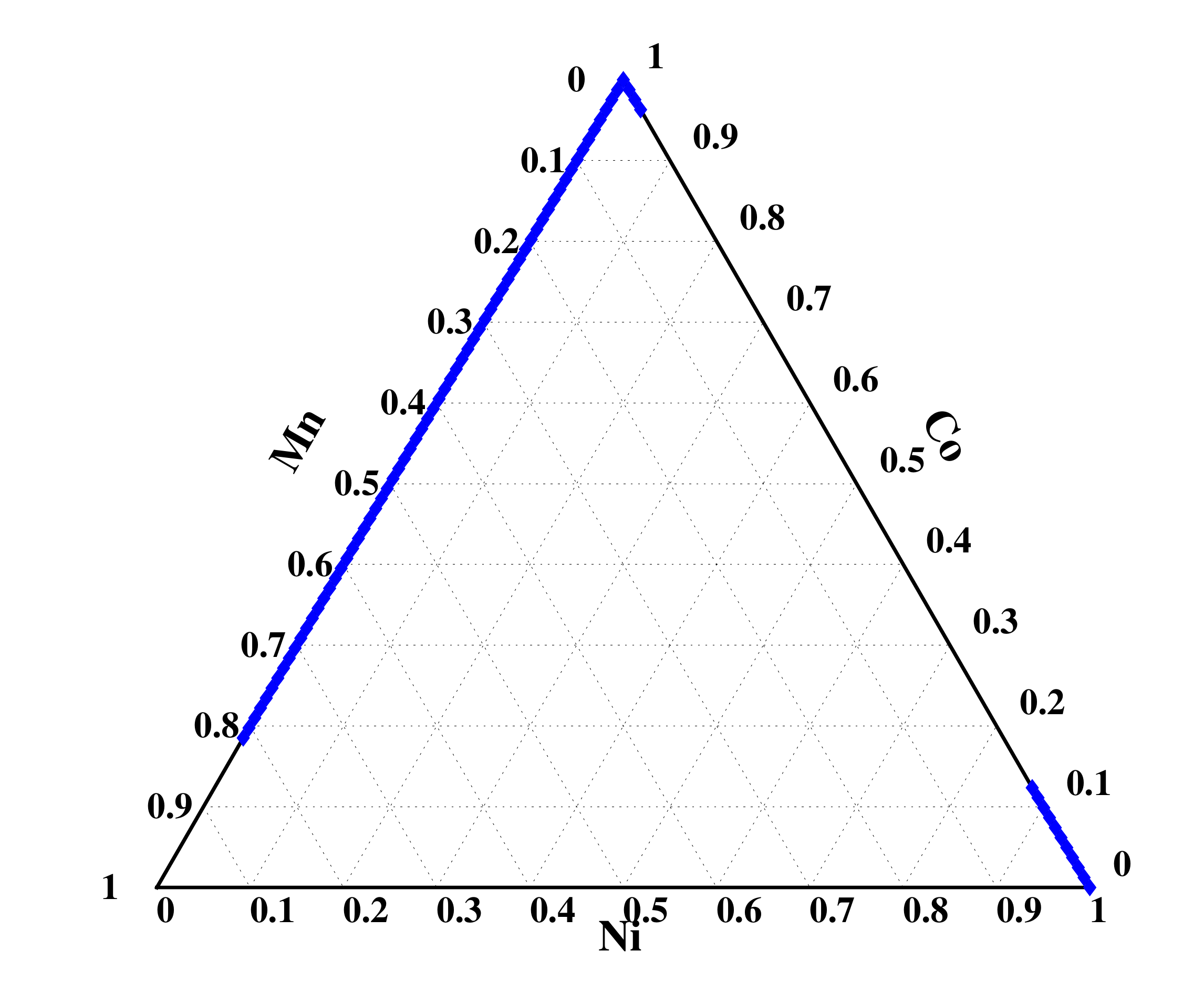}
				\put (-0.15,5) {\colorbox{white}{\large{\textbf{(a)}}}}
			\end{overpic}
			\hspace{5cm}
			\begin{overpic}[width=1.0\linewidth,trim={0 0cm 0 -0.2cm},clip]{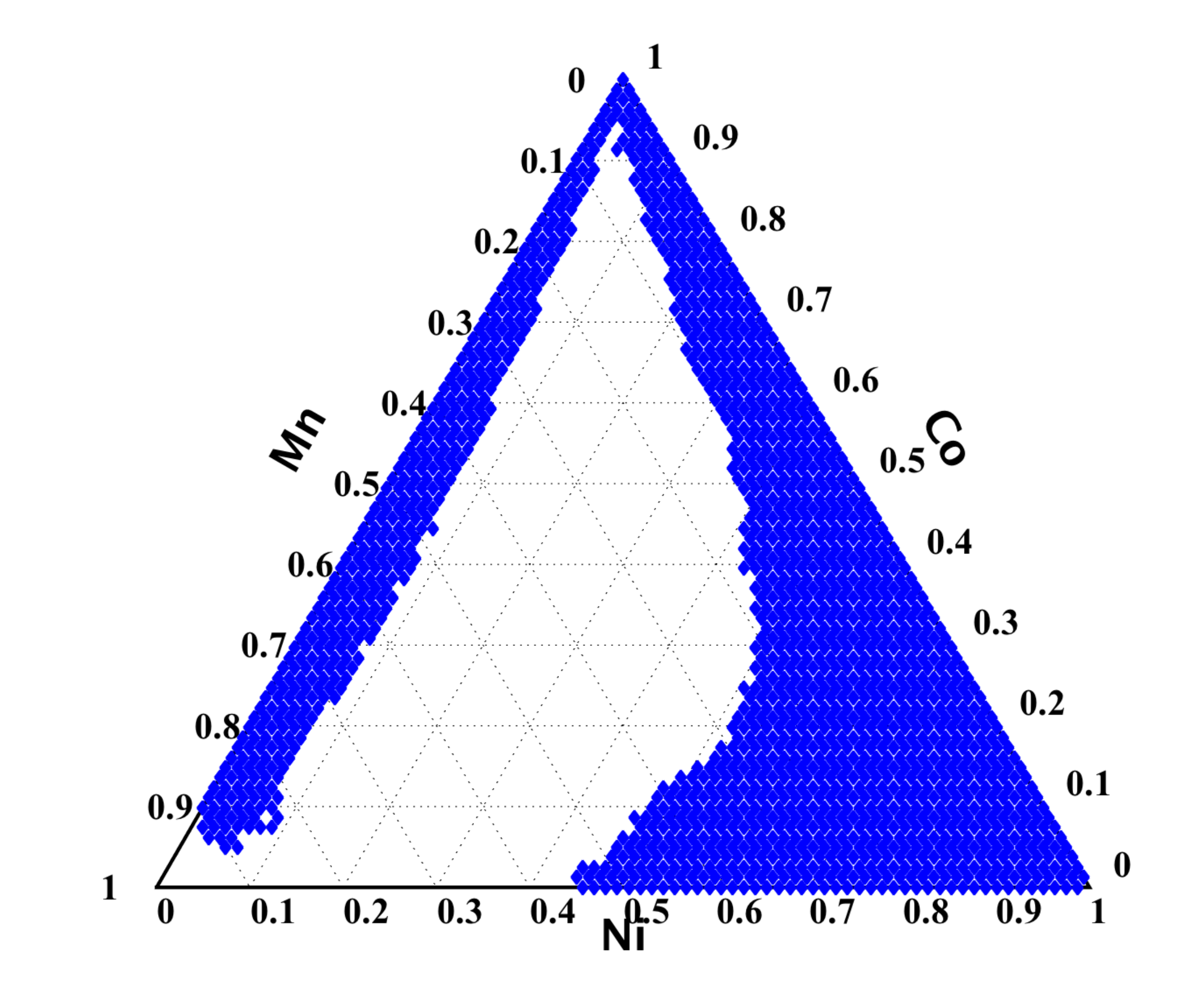}
				\put (0,4.5){\colorbox{white}{\large{\textbf{(c)}}}}
			\end{overpic}
		\end{minipage}
		\begin{minipage}[t]{0.475\linewidth}
		    \begin{overpic}[width= 1.0\linewidth,trim={0 0cm 0 0cm},clip]{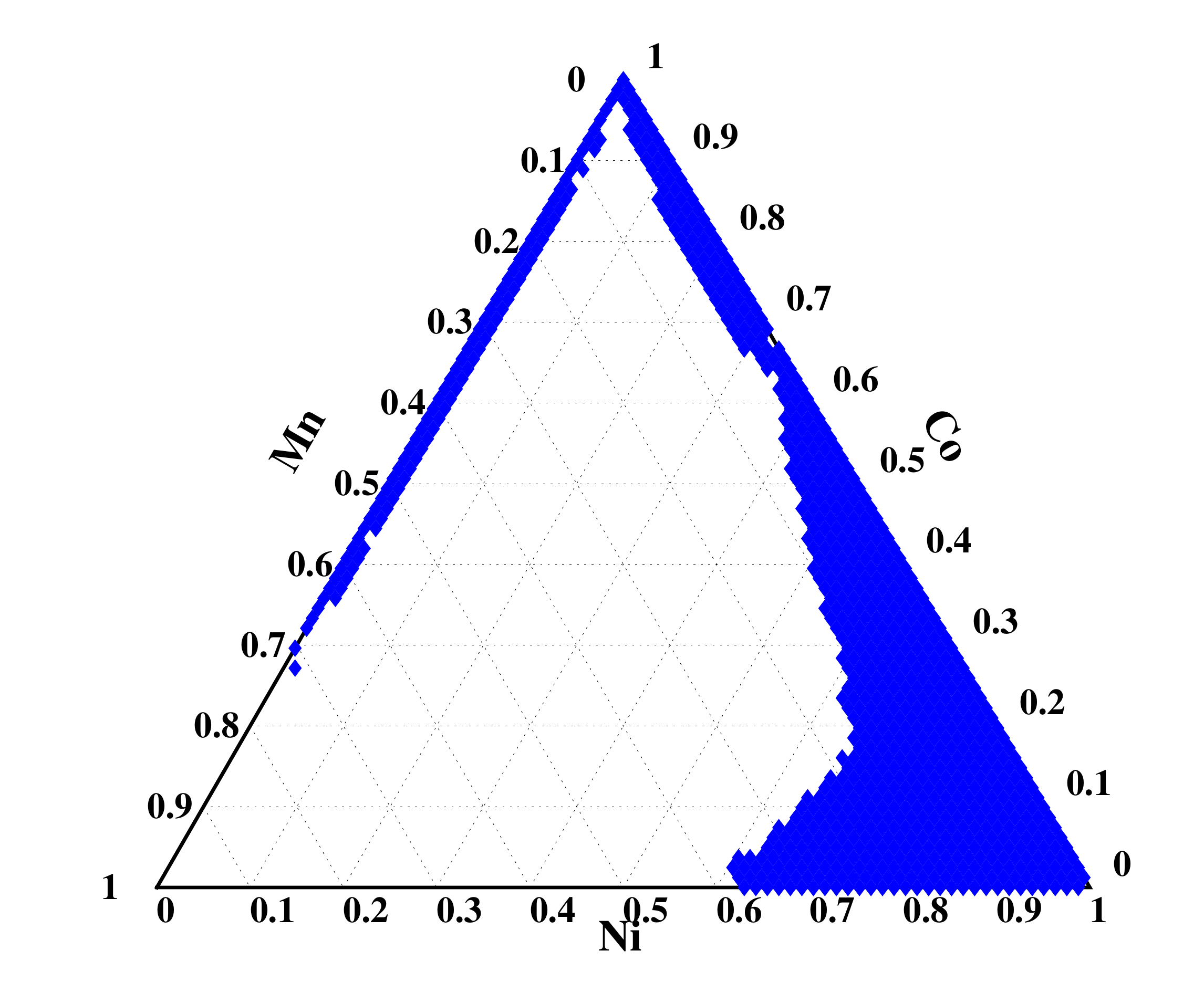}
				\put (0,5) {\colorbox{white}{\large{\textbf{(b)}}}}
			\end{overpic}
			\hspace{5cm}
			\begin{overpic}[width=1.0\linewidth,trim={0 0cm 0 -0.2cm},clip]{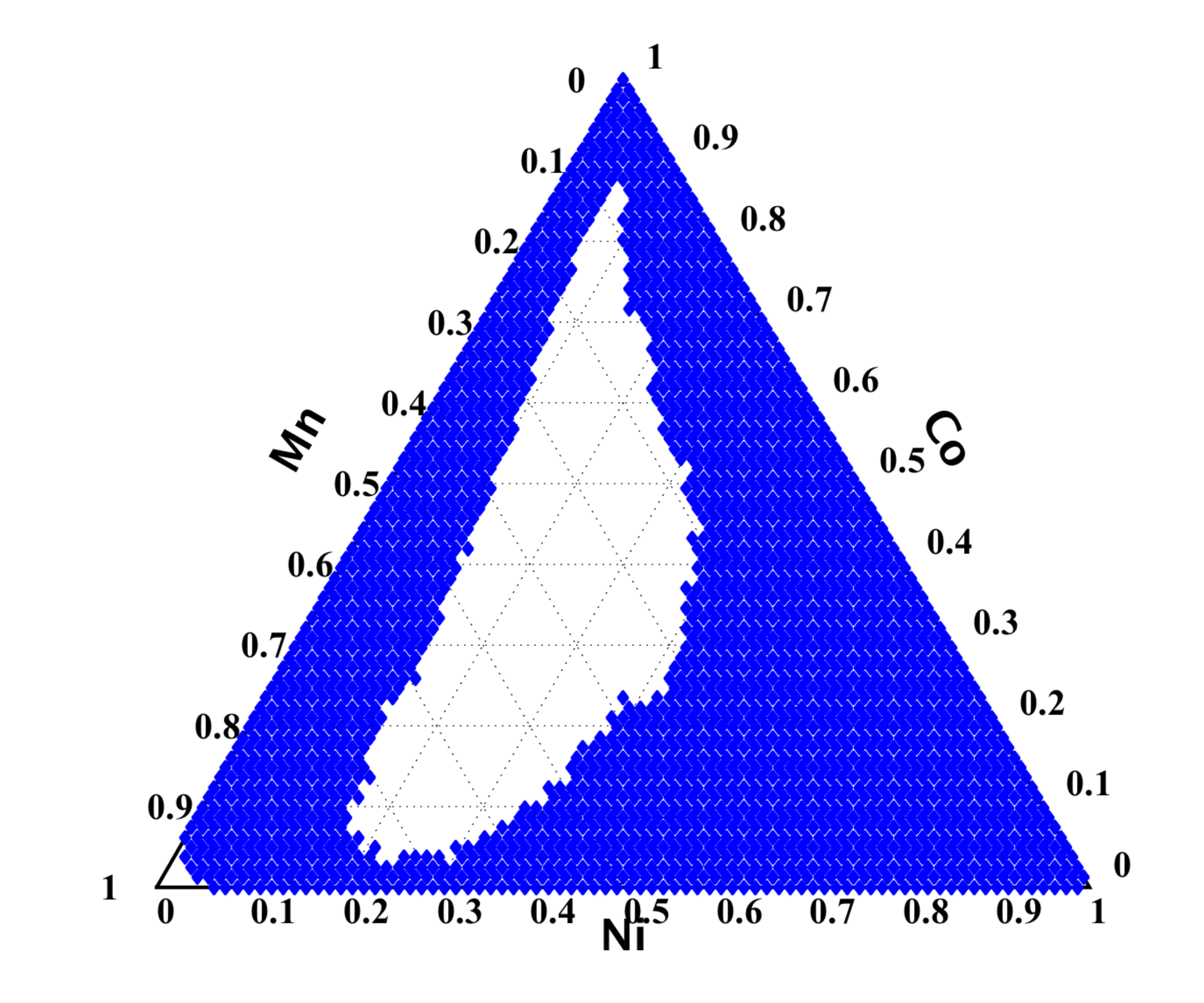}
				\put (0,4.5) {\colorbox{white}{\large{\textbf{(d)}}}}
			\end{overpic}
		\end{minipage}
		
		\caption{\label{fig:disorder} Predicted regions of disorder at (a) 298K (b) 900K, (c) 1300K and (d) 1700K.}
	\end{figure}

    Metropolis Monte Carlo simulations were then performed using all supercells ranging from 5x5 to 10x10 to sample a large collection of possible compositions and ordering to ultimately find the lowest energy states. Each of these supercells also contained 3 layers in the z direction to reflect the A-B-C layer stacking structure of these layered oxide materials. In order to aid in converging the lowest energy ground state, a simulated annealing scheme starting at 1500K and stepping down to 0K at 150K steps was performed. At each temperature in this scheme, the simulation was allowed to equilibrate a total number of steps equal to 100 times the number of lattice sites in the simulation box. Once the annealing is done, the simulation is then equilibrated at room temperature and then sampled for another total number of steps equal to 100 times the number of lattice sites in the simulation box. The average energy is then reported for that composition and for the purposes of understanding the ordered solid solution phases, the configurational entropy is ignored. This is justified as once the simulation was equilibrated at room temperature, there were little to no accepted transition metal swaps during any of the thermodynamic samplings and thus we treat the ordered phases as fully ordered with respect to the placement of transition metals. There was however significant spin configurational entropy as the simulations were performed above the magnetic phase transition temperature of NMC \cite{Chernova2007}. When the entropy was computed via thermodynamic integration of the heat capacity, it was within 10\% of the theoretical full spin entropy. The inclusion of full spin configurational entropy however, did not change the predictions for the ordered phases and therefore entropy was not explicitly computed for the ordered phases. The possibility of a disordered arrangement of transition metals was also considered. To simulate the disordered phase, a simulation was performed where all swaps were accepted and the energy of that configuration was sampled. This was done for 5000 steps and the average energy was reported. For the disordered phase, the full configurational entropy was included in the Gibbs energy and the results are compared to the ordered phase in Figure \ref{fig:disorder}. The blue region shows the range of compositions predicted to be a disordered solid solution at that temperature. It is seen, however, that most compositions at room temperature favor an ordered arrangement of transition metals, with the region of disorder growing as temperatures reach that of synthesis.  
    
    \section{Results}
\begin{figure}
		\includegraphics[width=.6\textwidth]{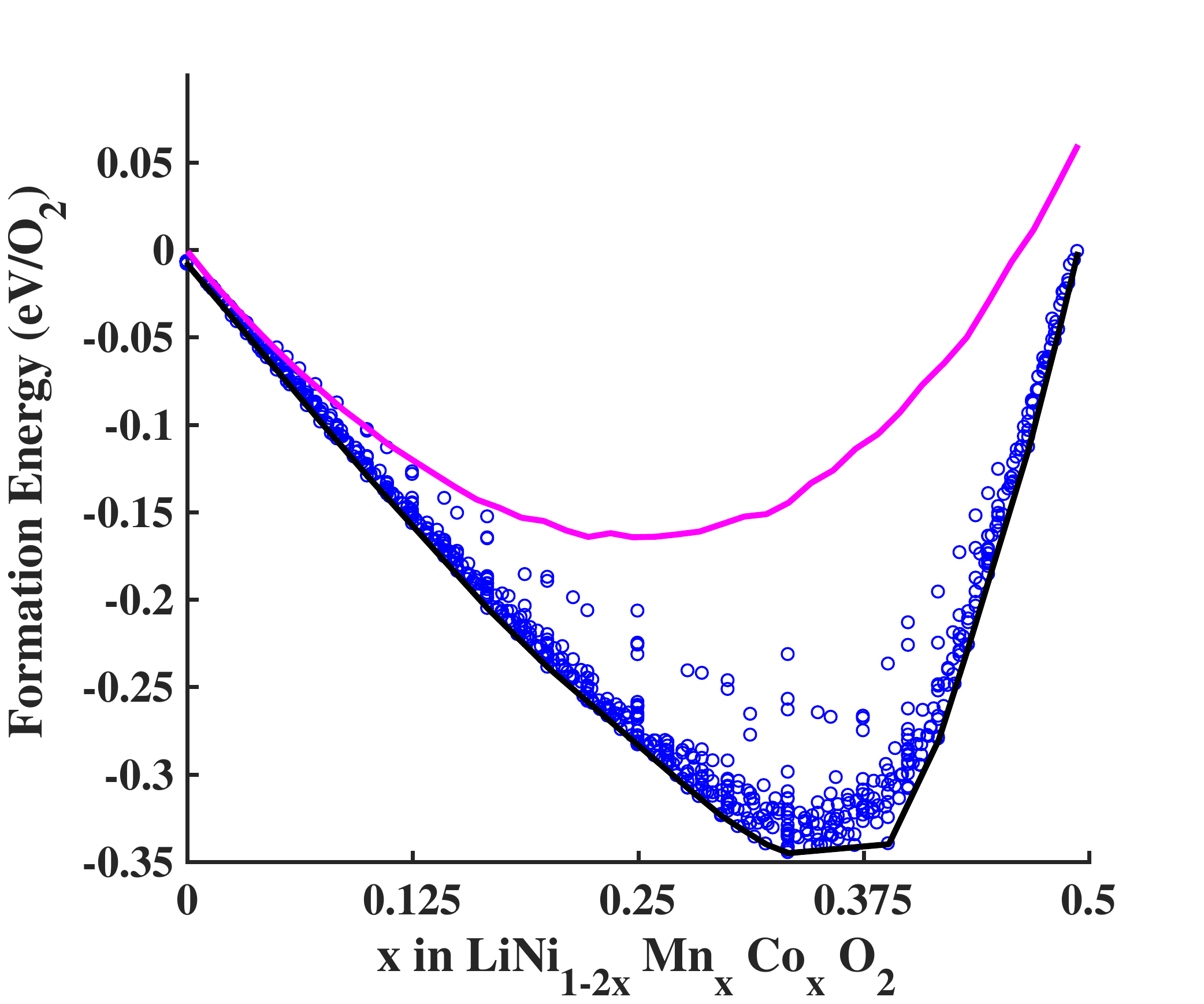}
		\caption{\label{slice} An example path in compositional phase space. The convex hull generated from the values given by the Monte Carlo simulations at 298K of the model from the self consistent DFT calculation is shown is black. The collection of blue dots are the results for various different Monte Carlo cells tested. Shown in pink is the prediction for the energy of the disordered phase at that composition for 298K.}
	\end{figure}

    To further understand the energetic differences between the ordered and disordered phases Figure \ref{slice} looks at a one dimensional slice along the compositional range LiNi$_{1-2x}$Mn$_x$Co$_x4$O$_2$. Within this range the ordered and disordered hulls were predicted for comparison. At approximately x=0.2 and above the energy difference between order and disorder is well above the errors associated with the calculations and therefore we predict with high confidence the materials with lower amounts of Ni to be ordered. In the high-Ni region, however, the energy difference is much smaller and could arguably be within error.

     As most of the phases a room temperature were predicted to be ordered, we then perform a hull analysis in an attempt to find the lowest-lying ordered compositional states and determine what compositional states will appear as pure states within a battery at room temperature. To do this, the points were fed into a 3D convex hull algorithm and all points on the hull with formation energy less than or equal to 0 eV were plotted as points on a ternary phase diagram. Any composition that does not appear as a vertex of the convex hull is thus predicted to exist as a phase separated mixture of neighboring compositions that do appear as a vertex. Additionally, we simulate the stability of these phases at room temperature, the expected operational temperature, rather than elevated synthesis temperatures as states that are stable at high temperature but metastable at room temperature will likely end up in the stable state as the cathode is cycled accompanied by volume change and capacity loss. The effect of higher temperature with regard to ordered and disorder is explore in Figure \ref{fig:disorder} for a collection of elevated temperatures typical of synthesis. The effect of higher temperature increases the entropic term and therefore effectively weaken the relative strength of the short range ordering. This leads to a larger region of compositions that exist as a disordered solid solutions.
  
	\begin{figure}
		\includegraphics[width=.48\textwidth]{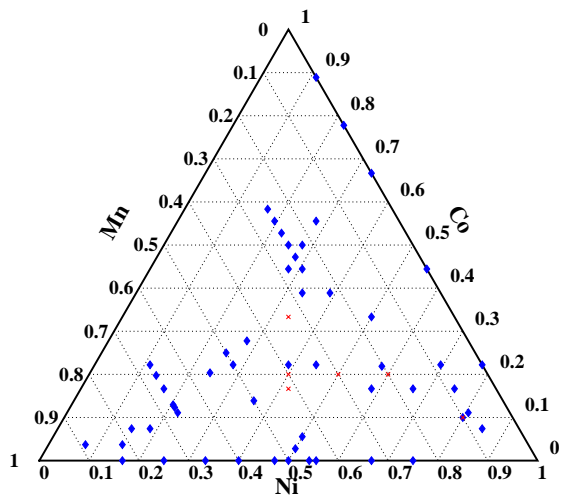}
		\caption{\label{single-xc} Predicted thermodynamically favorable compositional phases of NMC are shown as blue diamonds. All compositions not predicted to be thermodynamically stable are thus predicted to exist as a solution of neighboring stable phases. For reference, a collection of experimentally claimed phases from Ref. \citenum{Julien2016} are shown in red. A large disagreement between the the experimentally claimed phases and predicted phases should be noted, especially for the NMC111 phase.}
	\end{figure}
   
    As can be seen in Figure \ref{single-xc}, the predicted points in blue does not perfectly match the red points of experimentally used NMC phases only agreeing with a subset. Due to the sensitivity of the result on the choice of exchange correlation functional, among other sources of error, a careful treatment of uncertainty is required in order to understand if there is a reasonable prediction of new or slightly varying compositional compared to experiment.

    Although density functional theory is largely accepted as a computationally viable way to calculate the energy of a system from first principles, the topic of uncertainty quantification remains an issue. Attempts have been made to understand the uncertainty by testing predictions over large data sets and trying to understand the errors in statistical or regression models \cite{Pernot2018,DeWaele2016,Lejaeghere2016}. Other attempts have tried to incorporate the uncertainty into the generation of a new exchange correlation functionals \cite{Aldegunde2016,Petzold2012,Proppe2016,BEEF,mBEEF,Simm2016}. The exchange-correlation functional used within this work, BEEF-vdW, allows for the estimation of error that has been empirically fit to experiment. As discussed above, along with a main exchange-correlation functional that is used to self consistently converge the electron density and compute the energy, BEEF-vdW carries with it an adjustable number of functionals with slight perturbations from the main functional. The spread of these functionals has been pre-tuned to recreate a non-self consistent spread of energies that matches the error of the main functionals energetic prediction with respect to experimental training data. Previous work has utilized this spread of functionals to quantify the uncertainty at the level of GGA in the prediction of magnetic states\cite{c-value}, bulk properties\cite{ZeeshanPRB}, surface Pourbaix diagrams\cite{O-Pourbaix,Cl-Pourbaix}, scaling relations in oxygen reduction\cite{Yan2017,Deshpande2016}, formation enthalpies of reactions \cite{Rune2015CST}, and the prediction efficiency of DFT derived descriptors.\cite{DilipJPC}.

     \begin{figure}
		\includegraphics[width=.48\textwidth]{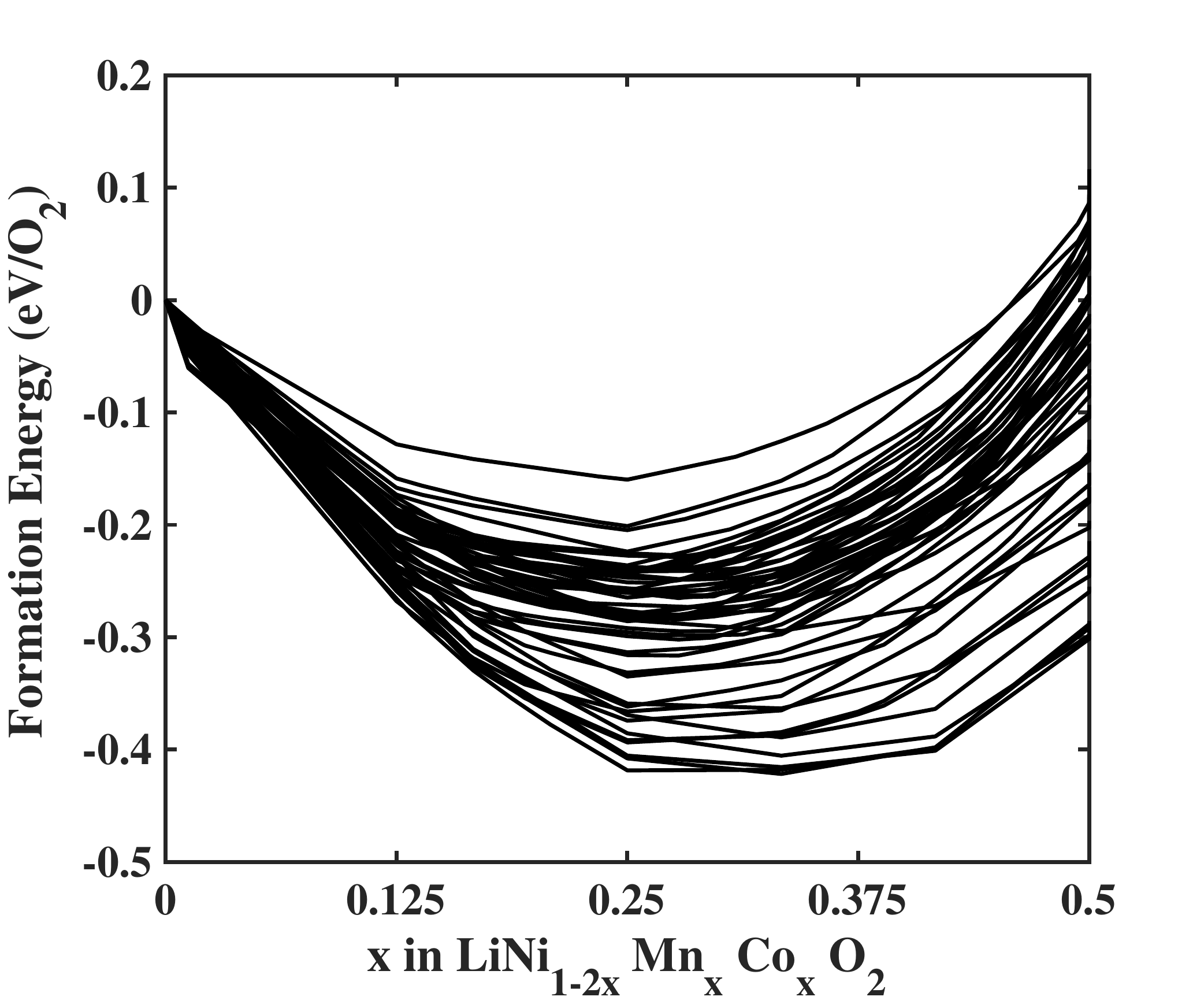}
		\caption{\label{ensemble_hulls} A plot of the the first 50 convex hulls from the 2000 total hulls.}
	\end{figure}  
	
     To further understand the challenge of propagating error through a hull analysis we again look at a single line within the ternary phase space. Figure \ref{ensemble_hulls} shows a collection of the fitted convex hulls over the compositional range of LiNi$_{1-2x}$Mn$_x$Co$_x$O$_2$. From this Figure, it is clear that the choice of functional largely affects the final predicted hull. The collection of hulls show a variety of not only energy predictions for the formation energies, but a variety of shapes. Therefore, different functionals will predict different phases to be stable and the DFT energy predictions are uncertain enough that the results of the phase diagram will change depending on your choice of functional.
     
\begin{figure}
		\includegraphics[width=.48\textwidth]{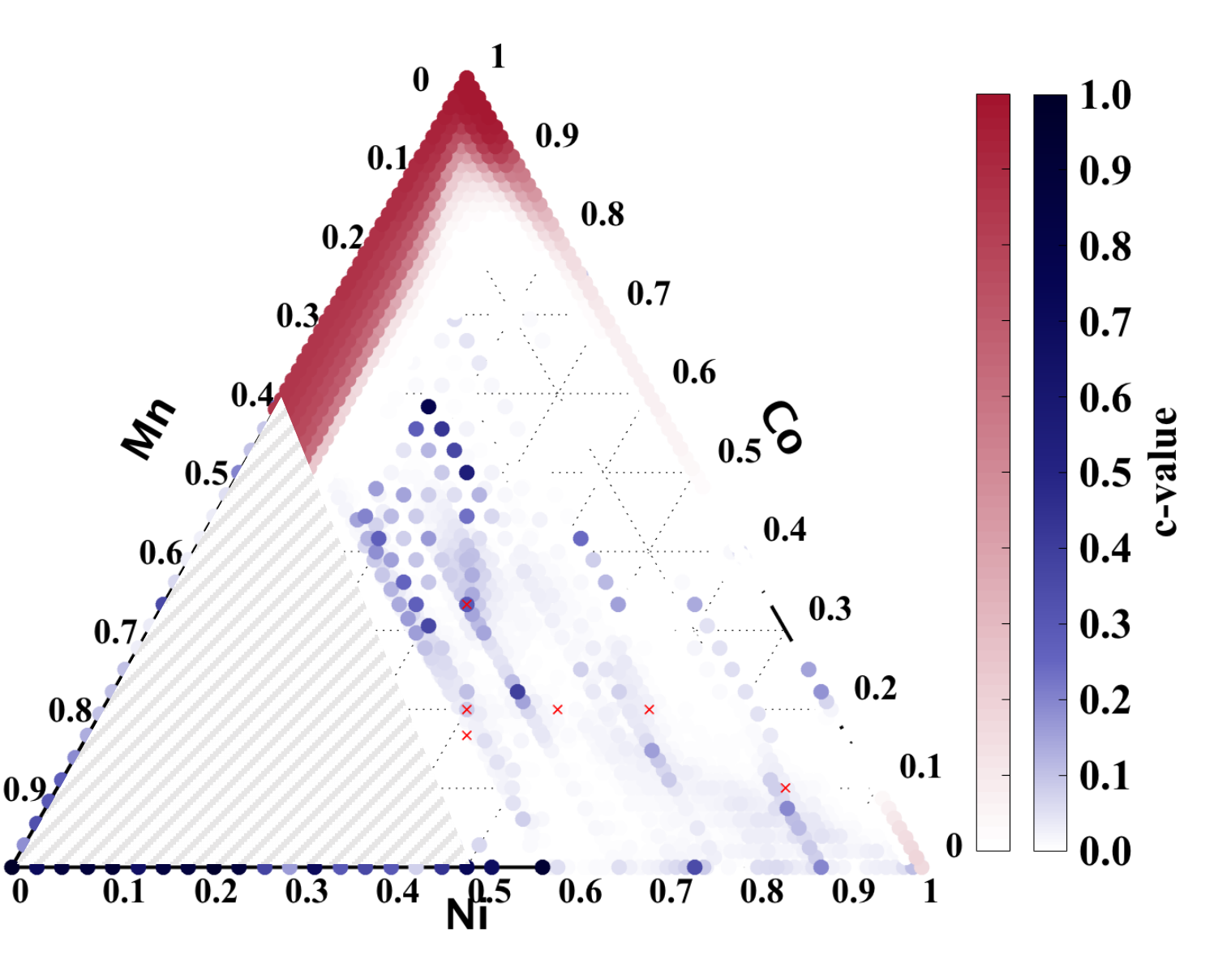}
		\caption{\label{combined-hull} A plot of the predominance of each composition on the 2000 different convex hulls fit at at a temperature of 298K in shown in blue. The darker the circles, the more often the phase appeared on the final hull generated by a single exchange correlation functionals and the more likely that it will appear as a pure compositional phase and not as a solution of other stable phases. A plot of the predicted disordered region 298K with predominance shown in dark red. Experimental phases from Ref. \citenum{Julien2016} are shown as bright red exes. The compositional region of high Mn that is known to be neither structurally stable over cycling has been blocked off in grey.}
	\end{figure}

     Additionally, as seen in Figure \ref{slice}, we have the largest uncertainty of whether the transition metals will be ordered or disordered at the high-Ni content region of interest. We therefore repeat our whole procedure of fitting the reduced order model, performing Monte Carlo at 298K to simulate thousands of compositions for the ordered and disordered phase, and feeding the results to a convex hull analysis for a set of 2000 ensembles generated from BEEF-vdW. We then count the number of times each specific composition appears on the convex hull as an ordered phase or as a disordered phase over all of the 2000 convex hulls ultimately generated. We then plot this normalized predominance in Figure \ref{combined-hull}.  This results in a collection of predicted phases near the experimentally used phases around which there is a smearing of uncertainty. We also see that with respect to exchange correlation functional, there is no uncertainty that the majority of the phases are ordered.

    \section{Analysis}
    The model used within this work assumes a fixed layered structure and hence, we limit our discussion to phases below a certain threshold Mn content. Beyond this level, the material will exhibit a layered-to-spinel structural transformation that is seen in the LiMnO$_2$ end member. Given recent work on the structural phase diagram of the Li-Ni-Mn-Co oxide pseudo-quaternary system, we set this cutoff to be around 40$\%$ Mn content at zero Ni content and approximately 50$\%$ at zero Co content \cite{Brown2015}. We also restrict our analysis to predicted phases that have a low Co content in hops of better satisfying the materials requirement surrounding the scarcity of Co. We show in Table \ref{tab:2} the 20 phases with the highest c-value that contain all three cations species and satisfy the criteria of less than 40$\%$ Co and less than 40$\%$ Mn for structural stability during cycling. Notice that the c-value for the NMC 111 phase, one of the most predominate phases in literature, is 0.304. We use this as a calibration for what level of confidence within our model has a plausible chance of occurring in reality, especially in regions where you see a smearing of probabilities around a specific composition as seen with 111. We notice this spread of compositions in areas near 811, between 442 and 532, and near 622. Additionally, the absence of the NMC111 phase in the prediction from only a single exchange correlation functional in Figure \ref{single-xc}, is one example of the importance of propagation of uncertainty in this work. Overall, we predict 29 phases that satisfy the design criteria with c-value above 0.1 which we consider to be of reasonable probability. There are 68 more phases satisfying our constraints with a c-value from 0.05 to 0.1. The full set of predicted phases, including compositions outside of our restricted design space, can be seen in Table S1 in the Supplementary Materials.

    As mentioned before, the 811 phase is a promising candidate for a high Ni-content cathode yet we do not recover this with high prediction confidence (c=0.011). We do however predict the phase 22 3 2 near this composition with c=0.198 as well as a smearing of phases immediately surrounding this phase with slightly smaller prediction confidence. Other experimentally claimed phases such as 622, 532 are also not seen in those exact proportions with significant confidence. In the case of 622, the phase predicted in those specific proportions with c=0.0255, while the nearest phase to this point NMC 17 6 4, appeared in the final result with a c-value of 0.1605. Finally of note is the predicted phase of interest lies between 532 and 442, with exact proportions of 432. We also see a few phases with promise in the region of ultra low Co content. We predict a disordered solid solution for the range of $LiNi_{1-x}Co_xO_2$ for $0<x<0.1$ as well as an ordered phase around x=0.2. We also predict some no Co phases of the form $LiNi_{1-x}Mn_xO_2$ for x=0.11,0.25,0.42 as well as recover the x=0.5 phase\cite{Arachi2005}.

    \begin{table}
		\caption{\label{tab:2} Predicted phases and their corresponding c-values. The exact proportions are given in the most reduced form as N M C}

			\begin{tabular}{rrrrrr}
				
				Ni & Mn & Co & c-value & \begin{tabular}{c}
				     Exact\\
				     Proportions
				\end{tabular} & Base  \\ \hline
				\\
				0.583 &
				0.417 &
				0.000&
				0.947&
				7 5 0&
				12
				\\
				
				0.500 &
				0.500 &
				0.000&
				0.7315&
				1 1 0&
				2
				\\
				
				0.444 &
				0.333 &
				0.222&
				0.3985&
				4 3 2 &
				9
				\\
			
				0.306 &
				0.389 &
				0.306 &
				0.3655 &
				11 14 11&
				36
				\\
				
				0.750 &
				0.250 &
				0.000 &
				0.3515&
				3 1 0 &
				4
				\\
			
				0.333 &
				0.333 &
				0.333 &
				0.304 &
				1 1 1&
				3
				\\
			
				0.278 &
				0.389 &
				0.333 &
				0.2745 &
				5 7 6&
				18
				\\
			
				0.250 &
				0.389 &
				0.361 &
    		    0.2595 &
    		    9 14 13&
    		    36
    		    \\
    		    
    		    0.889 &
				0.111 &
				0.000 &
    		    0.199 &
    		    8 1 0&
    		    9
    		    \\

    		    0.815 &
				0.111 &
				0.074 &
    		    0.198 &
    		    22 3 2&
    		    27
    		    \\
			
                0.200 &
                0.400 &
                0.400 &
                0.1845 &
                1 2 2&
				5
				\\
			
				0.778 &
				0.000 &
				0.222 &
				0.1805 &
				7 0 2&
				9
				\\
			
				0.630 &
				0.222 &
				0.128 &
				0.1605 &
				17 6 4&
				27
				\\
			
				0.222 &
				0.389 &
				0.389 &
				0.1585 &
				4 7 4&
				18
				\\
			
				0.500 &
				0.167 &
				0.333 &
				0.158 &
				3 1 2&
				6
				\\
			
				0.457 &
				0.333 &
				0.210 &
				0.154 &
				37 27 17&
				81
				\\
				
				0.750 &
				0.0 &
				0.25 &
				0.153 &
				3 0 1&
				4
				\\
				
				0.370 &
				0.333 &
				0.296 &
				0.1505 &
				10 9 8&
				27
				\\
				
				0.346 &
				0.333  &
				0.321 &
				0.15 &
				28 27 26 &
				81
				\\
				
				0.583 &
				0.083 &
				0.333 &
				0.142 &
				7 1 4 &
				12

			    \end{tabular}

	\end{table}

    The prediction of the exact compositions of NMC phases will allow more accurate predictions of the properties and degradation of these cathode materials. Further work using the specifically predicted phases here may allow better understanding of the tradeoffs of low Co cathodes and the limitations of the design space. While we do not present a full analysis of these phases, the work here provides a starting point to explore and understand the performance of these new phases and the possibility to achieve high performance ultra low-Co cathodes.

\begin{figure}
		\includegraphics[width=.48\textwidth]{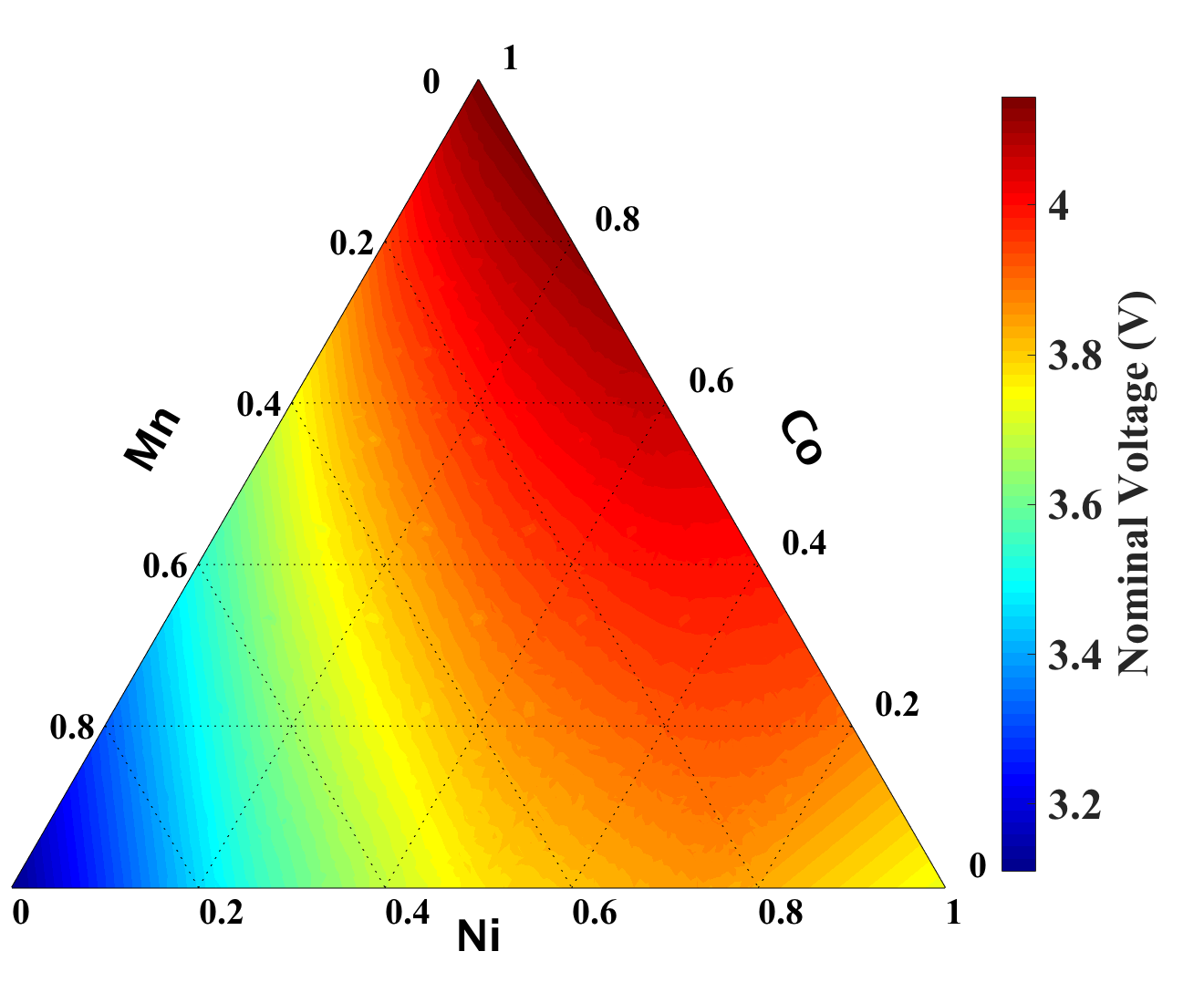}
		\caption{\label{volt} The predicted average voltage for all compositions calculated.}
	\end{figure}

    One of the easiest predictions that can be made once the Gibbs Free energy is known is the average voltage of the predicted NMC cathode with respect to Li/Li$^+$ given by the Nernst Equation.
    \[V = \dfrac{-\Delta G_f}{F}   \]
    The change in Gibbs energy is given by 
    \[ \Delta G= G_{\textrm{LiNMC}} - G_{\textrm{NMC}} - G_{\textrm{Li}}    \]
    If we assume that the delithiated phase has the same Gibbs energy as the delithiated end members as is discussed in the Supplementary Materials, we can calculate the average voltage of LiNi$_x$Mn$_y$Co$_z$O$_2$ as
    \[V = \dfrac{-\Delta G_{\textrm{LiNMC}}}{F} + xV_{\textrm{Ni}}+yV_{\textrm{Mn}}+zV_{\textrm{Co}}   \]
    A plot of the predicted average voltage over the entire phase can be seen in Figure \ref{volt} and a table of the predicted voltage of common NMC phases is presented in Table \ref{tab:3}. The prediction is compared to the predictions from AutoLion-1D \cite{AL1D}, a well benchmarked with respect to experiment model which allows for the simulation of an extremely slow discharge to better match the perfectly ideal average voltage predicted by this work. The results of this reproduces the benchmarked data to within 0.1 V for all NMC phases tested.
    
    \begin{table}
	\caption{\label{tab:3} The average voltage of various cathodes with respect to Li/Li$^+$ at 297K. To give an idea of the difference in the DFT predicted voltage and AL1D, the predicted average voltage of pure LiCoO$_2$ is also given. In general, the DFT predicted voltages are about 0.1eV or less higher than those predicted by AL1D. }

		\begin{tabular}{ccc}
			
			Phase & AL1D (V) & Model (V)\\ \hline
			\\
			111&
			3.83&
			3.88
			\\
			\\
			532&
			3.82&
			3.89
			\\
			\\
		 	811&
			3.83&
			3.86
			\\
			\\
			622&
			3.80&
			3.92
			\\
			\\
			001&
			3.98&
			4.10
		\end{tabular}

\end{table}
    
    With this prediction of voltage, we see there is little variation in the predicted nominal voltage of the cathode material as a function of transition metal content. Thus there in the context of average voltage, there is no tradeoff for low or no Co cathode materials. 
  
    It is well known that layered cathode materials with high Ni content present with measurable quantities of exchange between Ni and Li \cite{Zhang2010,Chen2016}. Within our analysis here, we have assumed the Li to remain in the alkali layer and the Ni to remain in the transition metal layer. To investigate this assumption, we have computed the formation energy, $E_f$, of a single defect of this kind and rough estimate of expected percent of cation mixing, $\eta$ at a given temperature to be 
    \[ p = \textrm{e}^{\frac{-\Delta E_f}{k_{B}T}}.  \]
    We see in Table \ref{tab:4} the predicted energies of these defects are rather large and predict extremely small quantities of these defects at both room temperature and even synthesis temperature as compared to some experimentally measured occurrences of this defect. We therefore assume that while this kind of defect is present in experimentally made cathodes, the largest driving force for the creation of the defects is unlikely to be thermodynamic but rather process related and the choice to ignore this effect within the main work is justified.

 \begin{table}
	\caption{\label{tab:4} The predicted formation energy in eV of the Li-Ni swapping defect for a collection of phases. }

		\begin{tabular}{ccccc}
			
			Phase & $E_f$ (eV) & $\eta_{298}$ (\%) & $\eta_{1200}$ (\%)& $\eta_{exp}$ (\%) \footnotemark[1]\\ \hline
			\\
			111&
			0.58&
			$2.0\times 10^{-7}$&
			0.69&
			1.56
			\\
			\\
			112&
			0.96&
			$5.6\times 10^{-15}$&
			$9.2 \times 10^{-3}$&
			0.70
			\\
			\\
		 	550&
			0.84&
			$6.1\times 10^{-13}$&
			0.029&
			6.7
			\\
			\\
			100&
			0.50&
			$3.3\times 10^{-7}$&
			0.78&
			
		\end{tabular}

	\footnotetext[1]{Ref \cite{Julien2016}}
\end{table}

\section{Conclusion}
    
    Given the recent circumstances around the availability and cost of Co, the high-Ni layered Li(Ni,Mn,Co,)O$_2$ phase space has received a renewed interest. Additionally, until now, no computational work involving the entire ternary phase space of NMC has investigated the thermodynamic favorability of the ordered versus disordered arrangement of transition metals within the lattice. We have presented a comprehensive exploration of this NMC phase space with the use of a reduced order model trained on DFT input and have predicted a large region of ordering despite the widespread treatment of this material as a disordered solid solution experimentally. The addition of uncertainty quantification related to the largest and least controllable source of error, the exchange correlation functional choice, leads to a more interpretability and reliable reduced order model and final prediction of compositional phases. We find that although there is large uncertainty in the specific ordered phases, there is less uncertainty with respect to exchange correlation functional for the prediction of ordering vs disorder. Additionally, for the ordered phases we see several collections of possible phases around what we believe to be the most likely phase within that region. With each of these regions near an experimentally widespread phase. From the understanding gained from uncertainty quantification of the reduced order model parameters and it's propagation to the final convex hull analysis, we were able to have a more comprehensive list of predicted phases and assign a confidence to these predictions. A collection of promising phases exist in the high Ni content region that satisfy both the constraint of low Mn content to avoid a structural transition to spinel during operation, as well the constraint of low Co content derived from economic concerns.

    This study lays out a starting point for future work in comprehensively studying these new phases both theoretically and experimentally to understand the tradeoffs of low-Co. Although we have demonstrated here a prediction of no significant decrease in operational potential, other effects such as capacity fade, open circuit voltage, thermal stability, and Li-Ni swapping are yet to be understood for these newly predicted phases. Additionally, the extension of this work to include other doping elements such as Al, and Mg may provide a larger design space to tune the properties of low-Co cathode materials and provide even more low-Co alternatives.

	\begin{acknowledgement}
		G. H. and V. V. acknowledge support from the Scott Institute for Energy Innovation at Carnegie Mellon and the National Science Foundation with award number CBET-1604898. G. H. would like to also thank Shashank Sripad for assistance with AL1D calculations.
	\end{acknowledgement}
	
	\bibliography{NMC}
	
	\clearpage



\section*{Supplementary material}

# Supporting Information for:

# Towards Ultra Low Cobalt Cathodes: A High Fidelity Computational Phase Search of Layered Li-Ni-Mn-Co Oxides

Gregory Houchins[†] and Venkatasubramanian Viswanathan[*,†]

†Department of Physics, Carnegie Mellon University, Pittsburgh, Pennsylvania 15213, USA

‡Wilton E. Scott Institute for Energy Innovation, Carnegie Mellon University, Pittsburgh, Pennsylvania 15213, USA

¶Department of Mechanical Engineering, Carnegie Mellon University, Pittsburgh, Pennsylvania 15213, USA

E-mail: venkvis@cmu.edu

## Delithated Case

The same method of preforming a set of density functional theory calculations as training data for a reduced order model was used in the case of a fully delithiated cathode $Ni_x$ $Mn_y$ $Co_{(1-x-y)}O_2$ which in the case of the Ni and Co end members is experimentally known to be in the O1 phase. This training data was fed into the same functional form of the reduced order model used for the lithiated case, and the fit model was simulated using Metropolis Monte Carlo. It was found, however that there are no mixed phases with a negative change in Gibbs energy. The resulting change in Gibbs Energy is very small and is therefore expected to the be kinetically stabilized especially if during operation, the cathode is not charged to

the fully delithiated state.

## Computational Details

Spin polarized density functional theory using the Generalized Gradient Approximation for the exchange-correlation energy were used to train the model for formation energy. The DFT training data was performed with various supercells of the O-3 structure of $LiMO_2$ where M=Ni,Mn,Co as is the experimentally seen phase for $LiNiO_2$ and $LiCoO_2$,[1,2] and is a metastable phase for $LiMnO_2$.[3?]

A 3x1, 2x2, 3x2, and 3x4 rhombohedral supercell as well as 1x1 and 2x1 triangular supercell with were used that contained 9 and 18 formula units respectively. For the 3x1, and 1x1 triangular cells all unique combinations of cation ordering within a single layer were used where the same cation ordering was repeated within the three layers. For all other supercells, a random selection of unique cation ordering each were used. Two possible 3x4 cation orderings for the a structure near the 811 phase (actual proportions of 10 1 1) were used to increase the model accuracy in the high-Ni region of the phase space. One extra calculation of the 532 phase in a 5x2 supercell from Wu el al.[4] was also include to assess the proposed cation ordering of a NMC532 phase. The initial spin state of the DFT training data was randomly chosen to be positive or negative to create a random sampling of spin interaction and in most cases the final energy converged to the same spin state as was initialized. The final spin state, however, was always used in generating the cluster terms.

All density functional calculations were performed using the Projector Augmented Wavefunction as implement in realspace within the GPAW software[5] with a grid spacing of 0.16 Åand [12,12,2] k-points per conventional cell, with Monkhorst-Pack grid. Both the grid spacing and the k-point mesh were converged to $10^{-4}$eV with respect to the predicted formation energy of test LiNMC mixtures. Within each DFT calculation with these grid-point and k-point specifications, the densities were converged to 0.0001 electrons, and the energies were converged to $10^{-4}$ eV per formula unit. The specific lattice parameters of each cation com-

position and ordering was found by varying the lattice parameters in the a and b directions equally with strain of $x = \dfrac{a}{a_0} = 0.9, 0.95, 1, 1.05$, and $1.1$ where $a = b$ are the experimental lattice parameters of $LiNiO_2$, while fixing the c direction. The volumes and energies of at each point were fit to a jellium equation of state[6] and the ideal lattice parameters were extracted. This process was then repeated by fixing the a and b constants and varying the c constants then fitting to find the ideal constant in the way as before. The resulting structure at the optimized lattice parameters was then relaxed to a maximum force of 0.01eV/ Å.

## Reduced order Model

It should be noted that the inclusion of all n-body terms in the model would lead to a rank deficient system of equation when trying to fit the interaction coefficients through a system of linear equations. This is because there is an external constraints of all the lattice sites being filled that creates a linear dependence on the number of cations with the number of nearest neighbor interactions, next nearest neighbor interaction, etc. Because of this, we choose to artificially set the all homogenous n-body interactions to be zero. What is left in the model is the relative favorability of a given interaction compared to all other interactions. The model was fit using multivariate normal regression where the design matrix of number of interactions and the formation energies were normalized to one stoichiometric unit of $LiMO_2$. This prevents overweighting the larger supercells. The magnitude of the spins of each cation is assumed to be constant for all composition of Ni, Mn, and Co with the same level of lithiation. This assumption allows for the reduced order model to be solved more easily since allowing the magnetic moments of each atom to vary continuously and unbounded during Monte Carlo simulations would cause the spin interaction term to be unbounded.

Once the final model was chosen, the convergence of the model as a function of number of training points was tested with leave one out cross validation as seen in Figure S1. The root mean square error levels of at around 100 training points and appears constant at the full 120 point data set.

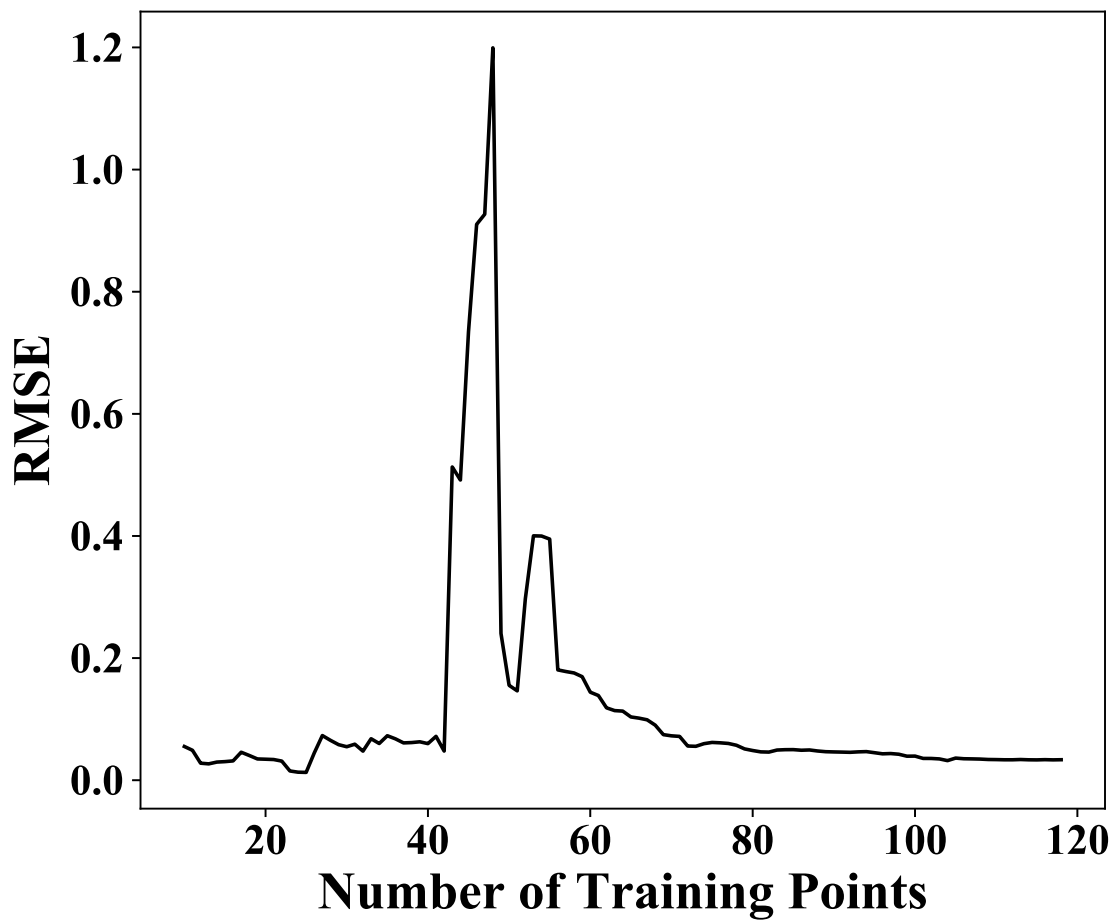

Figure S1: The root mean squared error (RMSE) from leave one out cross validation as a function of number of training parameters. A random subset of 40 DFT calculations was chosen and one by one a new data point was added. For each subset, leave one out cross validation was performed.

## Monte Carlo Simulations

Once the model is trained, the enthalpy of formation at finite temperature is simulated using a Metropolis Monte Carlo algorithm. For each composition of transition metals, a random initial cation and spin ordering is assumed. Then randomly the spin of a single cation or the position of two cations is swapped. After each swap, the condition for acceptance is given by estimating the probability of the swap happening spontaneously.

$$P_{1 \to 2} = \mathrm{e}^{\frac{-E}{k_B T}}$$

Where $\Delta E$ is the energy change of the swap. If the energy change is negative then the swap is always kept. If the energy change is positive, it is kept if the probability is less than than a randomly generated number between 0 and 1. This algorithm allows for the thermal population of states and given a sufficiently long simulation, will accurately predict the probability of thermal population.

Each simulation is thermalized via simulated annealing starting at 1500K and then decreasing by steps of 150K to 0K with each step simulated over 100 times the number of lattice sites. The enthalpy is then calculated to be the average enthalpy of all of the sweeps. This procedure can be seen for an example case in Figure S2.

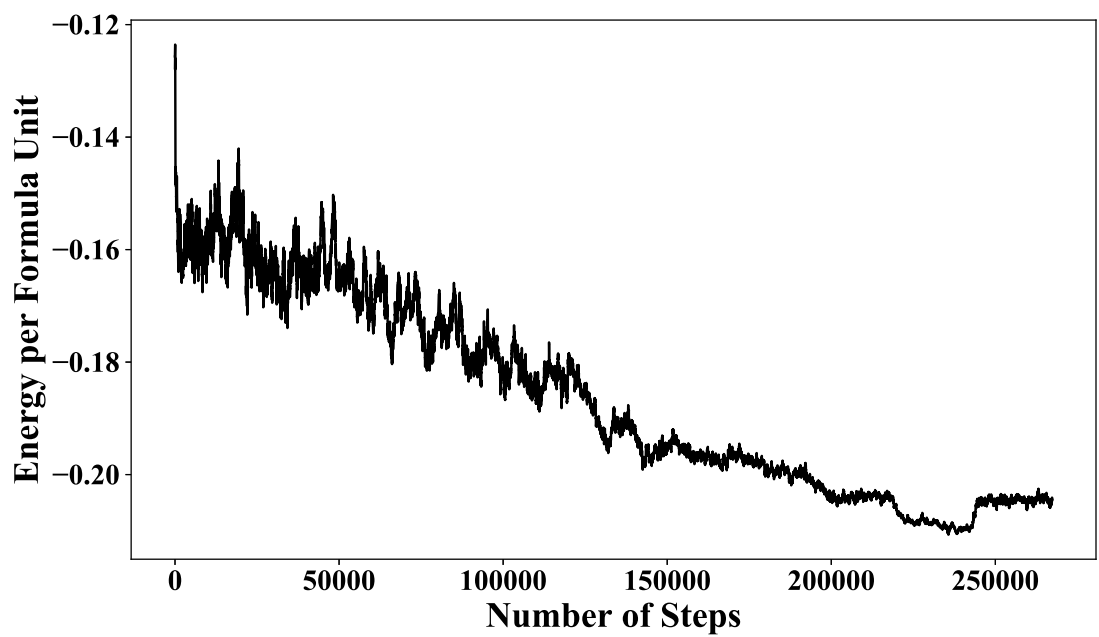

Figure S2: A plot of the energy per formula unit as a function of step number during the simulated annealing step followed but the 100 steps per number of lattice site thermalization step at 298K. The simulation is for a 9x9 supercell to demonstrate the convergence of a larger supercells.



\end{document}